\documentstyle[prl,aps,amssymb]{revtex}
\input{epsf}
\newcommand{\beq}{\begin{equation}}
\newcommand{\eneq}{\end{equation}}
\begin{document}

\tolerance 10000

\twocolumn[\hsize\textwidth\columnwidth\hsize\csname %
@twocolumnfalse\endcsname

\draft

\title{Spinon Attraction in  Spin-1/2 Antiferromagnetic Chains }

\author {B. A. Bernevig, D. Giuliano and R. B. Laughlin}

\address{Department of Physics, Stanford University,
        Stanford, California 94305}

%\twocolumn[
\date{\today}
\maketitle
\widetext

\begin{abstract}
%\vspace*{-1.0truecm}
\begin{center}

\parbox{14cm}{We derive the representation of the two-spinon wavefunction
for the Haldane-Shastry model in terms of the spinon coordinates. This
result allows us to rigorously analyze spinon interaction and its physical
effects. We show that spinons attract one another.  The attraction gets
stronger as the size of the system is increased and, in the thermodynamic
limit, determines the power law with which the susceptibility diverges.}

\end{center}
\end{abstract}

\pacs{
\hspace{1.9cm}
PACS numbers: {75.10.Jm, 75.40.Gb, 05.30.Pr}
}
]

\narrowtext

Interacting spin-1/2 antiferromagnetic spin chains in 1 dimension 
 exhibit low-lying excitations carrying spin-1/2, called spinons
\cite{fadeev}. The Brillouin zone for one spinon is halved
\cite{fadeev,haldane,shastry}, and spinons are semions, i.e., particles
with statistics half that of regular fermions \cite{spinongas,frank}.  The
large-scale physics of a generic 1-d antiferromagnet with short-range
interaction is given by a spinon gas \cite{spinongas}. The corresponding
energy for an $N$-spinon solution is the sum of the energies of each
isolated spinon, plus corrections that go to zero in the thermodynamic
limit.

The additivity of the energy is usually claimed as an evidence for a
spinon gas to be an ensemble of free semions \cite{hald3,hazi,talstra}. 
In this letter we challenge this idea by carefully analyzing the
interaction between spinons in an exact solution of a particular 1-d
antiferromagnet: the Haldane-Shastry model (HSM). The HSM is a system of
spins on a circular lattice interacting via an antiferromagnetic
interaction inversely proportional to the square of the chord between the
corresponding sites. The Hamiltonian is given by

\beq
{\cal H}_{HS} = J ( \frac{2 \pi}{ N } )^2 \sum_{ \alpha < \beta }^N
\frac{ \vec{S}_\alpha \cdot \vec{S}_\beta}{ | z_\alpha - z_\beta |^2} 
\; \; \; ,
\label{hsmodel}
\eneq

\noindent
where $z_\alpha$=$\exp (2 \pi i \alpha /N)$ and $\alpha$ is the lattice
site The HSM is the simplest exactly-solvable interacting antiferromagnet
in 1-d.  It is the prototype of a 1-d spinon gas since it does not take
marginal logarithmic corrections, in contrast, for instance, with the
behavior of the Heisenberg model \cite{haldane,shastry}.

Many-spinon solutions of the HSM have been constructed \cite{haldane} in
analogy to the corresponding spinless continuum version of the model
\cite{sutherland}.  However, from the corresponding ``plane wave''
representation of the many-spinon wavefunction, the persistence of a
spinon interaction in the thermodynamic limit is not at all transparent.
Indeed, in the thermodynamic limit the energy is the sum of the energies
of each isolated spinon. However, the interaction between spinons is
``hidden'' in the nontrivial relation between the canonical momenta, which
label the states, and the kinetic momenta, which determine the energy
\cite{frank}.

\begin{figure}
\epsfbox{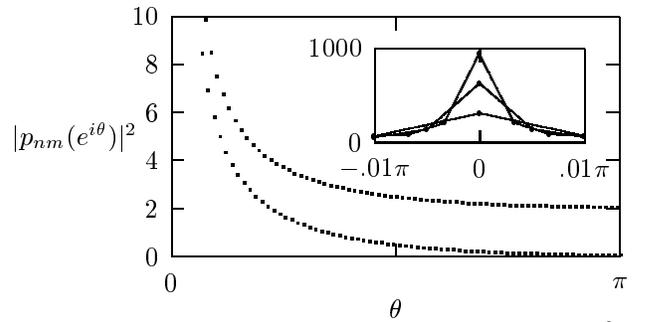}
\caption{Square of two-spinon wavefunction $|p_{mn}(z)|^2$ defined by Eq.
        (\ref{bigboy2}) for the case of $N=300$, $m=N/2 - 1$, and $n = 0$.  
	At large
        separations the probability oscillates between 0 and 2 and
        averages to 1.  The inset shows this function close to
        the origin for $N$ =  200, 400, and 600.  The value at the
        origin diverges in the thermodynamic limit.}
\label{fig1}
\end{figure}

In this paper we work out the real-space coordinate representation for
two-spinon eigenstates of ${\cal H}_{HS}$ and the corresponding
Schr\"odinger equation. Spinon interaction and its nature follow
straightforwardly from the behavior of the exact solution of this
equation. In Fig.\ref{fig1} we plot the result. While at large separations
the probability amplitude is independent of spinon separation, as it is
appropriate for noninteracting particles, at short separations there is a
huge enhancement.  Such a resonant enhancement is a clear evidence for a
short range, attractive interaction between spinons. As we show in
Fig.\ref{fig1}, this enhancement gets sharpened as the number of
sites increases, at odds with the belief that spinon interaction and its
effects disappear in the thermodynamic limit. 

Spinon dynamics determine the low-energy physics of the HSM. 1-d
interacting antiferromagnets do not order and, accordingly, the spin-1
spin-wave (SW) is an unstable excitation of the HSM. The SW is absolutely
unstable at any energy and momentum against decay into a spinon pair
\cite{fadeev}. This causes non-analyticities in the SW propagator, the
dynamical spin susceptibility (DSS) $\chi_q ( \omega )$. $\chi_q ( \omega
)$ develops a branch cut at the threshold energy for a SW and a broad
continuum above this threshold. Broad spectra have been observed by means
of neutron scattering on quasi 1-d samples \cite{tennant}, which
experimentally substantiates this scenario. However, the continuum is not
flat, as would be the case is it were a spinon joint density of states,
but rather has a divergent square root edge. We shall show that it is the
spinon interaction which makes the matrix element for the decay of the
spin wave into spinon pairs huge at threshold, and causes this divergence.  
We explicitly prove that, in the thermodynamic limit, the spinon
attraction turns into the square root divergence in the DSS.  Spinon
interaction and its relation to the DSS are the main result of our work.

Let us begin with some basic results from the HSM.  In the even-$N$ case
the ground state of ${\cal H}_{HS}$ (eq.(\ref{hsmodel}))  is a disordered
spin singlet, whose wavefunction is given by

\beq
\Psi_{GS} ( z_1 , \ldots , z_M ) = \prod_{i<j}^M ( z_i - z_j)^2
\prod_j^M  z_j \; \; \; ,
\eneq

\noindent
where $M=N/2$ and the $\{j\}$'s denote the positions of $\uparrow$-spins,
all the others being $\downarrow$. The corresponding energy is given by
$E_{GS} = - J ( \pi^2 / 24 ) ( N + 5/N)$ \cite{haldane,shastry,chia}. 
Elementary excitations above $\Psi_{GS}$ are spinons--spin-1/2 defects in
the otherwise featureless disordered sea. A $\downarrow$ spinon localized
at $\alpha$ can be thought of as a singlet sea where the spin at $\alpha$
is constrained to be $\downarrow$.  The corresponding wavefunction is

\beq
\Psi_\alpha (z_1 , \ldots, z_M) = \prod_j^M ( z_\alpha - z_j ) 
\prod_{i<j}^M ( z_i - z_j)^2 \prod_j^M z_j \; \; \; ,
\eneq

\noindent
where now $N$ is odd and $M=(N-1)/2$. A one-spinon eigenstate of ${\cal
H}_{HS}$ is constructed by making the plane-wave superposition

\beq
\Psi_m ( z_1 , \ldots , z_M ) = \frac{1}{N} \sum_{ \alpha = 1}^N (
z_\alpha^*
)^m \Psi_\alpha ( z_1 , \ldots , z_M) \; \; \; .
\label{equ22}
\eneq

\noindent
The corresponding energy is 

\beq
E_m = -J \frac{\pi^2}{24}(N - \frac{1}{N} ) + \frac{J}{2}
(\frac{2\pi}{N})^2 m ( M - m) \; \; \; .
\eneq

\noindent
$\Psi_m$ has also a well-defined crystal momentum: $q_m = ( \pi / 2) N - (
2 \pi / N ) ( m + 1 /4 )$ (mod $2\pi$). In terms of $q_m$ the energy with
respect to the ground state is $E (q_m) = ( J/2 ) [ (\pi / 2 )^2 - q_m^2
]$ (mod $\pi$).

Spinons do not lose their identity when many of them are present. $L$
spinons can be thought of as a disordered sea with the spin at $L$ sites
constrained to be $\downarrow$ \cite{haldane}. For two spinons this means
that the corresponding wavefunction for a pair of localized spinons at
$\alpha$ and $\beta$ is given by ($M=N/2-1$)

\[
\Psi_{ \alpha \beta} ( z_1 , \ldots , z_M) = 
\]

\beq
\prod_{j}^M ( z_\alpha - z_j )  ( z_\beta - z_j ) 
\prod_{i<j}^M ( z_i - z_j )^2 \prod_j^M z_j
\; \; \; .
\label{equ3}
\eneq

\noindent
$\Psi_{\alpha \beta}$ can be analytically extended to any value of
$z_\alpha, z_\beta$ on the unit circle. As $z_\alpha,z_\beta$ are lattice
sites, they are interpreted as locations of $\downarrow$-spins. 

States with two spinons carrying well-defined crystal momentum are given
by the lattice plane waves which have the expression

\beq
\Psi_{ m n  } ( z_1 , \ldots , z_M ) =
\sum_{\alpha, \beta}^N 
\frac{(z_\alpha^*)^m (z_\beta^*)^n}{N^2}
\Psi_{ \alpha \beta } ( z_1 , \ldots , z_M ) \; .
\eneq

\noindent
The total crystal momentum of $\Psi_{mn}$ is $q = ( \pi/ 2) (N - 2) + q_m
+ q_n$ (mod $2 \pi$)  and $q_m,q_n$ are the momenta of each spinon. The
$\Psi_{mn}$ are an overcomplete set. A set of linearly independent states
is constructed by taking only the $\Psi_{ mn}$ with $M \geq m \geq n
\geq 0$.  Two-spinon energy eigenstates are linear superpositions of
these:

\beq
\Phi_{mn} = \sum_{l=0}^{\ell_M} a^{mn}_\ell
\Psi_{ m + \ell , n-\ell} 
\; \; \; ,
\label{tras1}
\eneq

\noindent
where $\ell_M = n$ if $m+n<M$, $\ell_M=M-m$ otherwise. The coefficients
$a^{mn}_\ell$ are \cite{sutherland,chia}

\begin{equation}
a_{\ell}^{mn} = -  \frac{ ( m - n + 2 \ell  ) \;}
{ 2 \ell ( \ell + m -n + \frac{1}{2} )}
\sum_{ k=1}^\ell a^{mn}_{ k - 1}
\; \; \; \; \;    (a_0 = 1)
\label{fmn}
\end{equation}

\noindent
and the corresponding eigenvalue is

\begin{displaymath}
E_{mn} = - J ( \frac{\pi^2}{24} ) (  N + \frac{5}{N} ) +
\end{displaymath}

\beq
[ E (q_m ) + E ( q_n) - \frac{ \pi J}{N} \frac{|q_m - q_n|}{2} ]
 \; \; \;  (q_m\leq q_n) \; \; . 
\label{2spen}
\eneq

\noindent
$E_{mn}$ is the sum of the ground-state contribution, $E_{GS} = - J (
\pi^2 / 24 ) ( N + 5 / N )$, and $E (q_m , q_n)$, which is the two-spinon
energy above the ground state. $E (q_m , q_n)$ is the sum of the energies
of two isolated spinons plus a negative interaction contribution that
becomes negligibly small thermodynamic limit.

The norm of $\Phi_{mn}$ can be computed by means of a recursive
procedure, based on the operator $e_1 ( z_1 , \ldots , z_M ) = z_1 +
\ldots + z_M$.  For any wavefunction of the form $\Phi \times \Psi_{GS}$,
where $\Phi$ is a symmetric polynomial, we have

\begin{displaymath}
{\cal H} \Phi  \Psi_{GS} = E_{GS} \Phi \Psi_{GS} +
\frac{J}{2} (\frac{2\pi}{N})^2 \Psi_{GS} \biggl\{ \frac{1}{2} \biggl[ 
\sum_j z_j^2 \frac{\partial^2}{\partial z_j^2}
\end{displaymath} 

\begin{equation}
+ 4 \sum_{j \neq k}
\frac{z_j^2}{z_j - z_k} \frac{\partial}{\partial z_j} \biggr] - 
\frac{N - 3}{2} 
\sum_j z_j
\frac{\partial}{\partial z_j} \biggr\} \Phi
\; \; \; ,
\end{equation}

\noindent
and thus

\begin{displaymath}
{\cal H} e_1 \Phi \Psi_{GS} - e_1 {\cal H} \Phi \Psi_{GS} =
\frac{J}{2} (\frac{2\pi}{N})^2 
\end{displaymath}

\begin{equation}
\times \Psi_{GS} \biggl[ \sum_j z_j^2 \frac{\partial}{\partial z_j}
+ \frac{N-3}{2} e_1 \biggr] \Phi \; \; \; .
\end{equation}

\noindent
>From the matrix elements of the commutator between ${\cal H}_{HS}$ and
$e_1$ under the inner product $\langle f | g \rangle = \sum_{z_1 , \ldots
, z_M } f^* ( z_1 , \ldots , z_M ) g ( z_1 , \ldots , z_M)$
we find that, for the two spinon eigenstates ($M=N/2 -1$)

\beq
\frac{ \langle \Phi_{m-1 , n} |  e_1 | \Phi_{mn}  \rangle}{ \langle
\Phi_{m-1 , n } | \Phi_{m-1 , n} \rangle} =
\frac{ - ( M - m +1) }{ 2 ( M - m +\frac{1}{2}) } 
\eneq

\beq
\frac{  \langle \Phi_{m-1 , n} |  e_1 | \Phi_{mn}  \rangle}{
\langle \Phi_{ m n } | \Phi_{ m n } \rangle }
= \frac{ - ( m + \frac{1}{2} ) ( m - n)^2 }{ 2 ( m - n + \frac{1}{2} ) 
m ( m - n - \frac{1}{2} ) }
\eneq

\beq
\frac{
\langle \Phi_{ m , n-1} |  e_1 | \Phi_{ m n}  \rangle}{
\langle \Phi_{ mn} | \Phi_{ mn} \rangle } = 
 \frac{ - n}{ 2 ( n - \frac{1}{2} )} 
\eneq

\begin{displaymath}
\frac{ \langle \Phi_{ m , n-1} |  e_1 | \Phi_{ m n} \rangle  }{
 \langle \Phi_{ m , n-1} | \Phi_{ m , n-1 } \rangle } =
\end{displaymath}

\beq 
 \frac{ - ( M - n + \frac{3}{2} ) ( m - n + 1 )^2}{ 2 ( m - n +
\frac{3}{2}
) ( m - n + \frac{1}{2} ) ( M - n + 1 )}
\; \; \; .
\eneq

\noindent
Combining these expressions, one then finds by induction that

\begin{displaymath}
\frac{\langle \Phi_{mn} | \Phi_{mn} \rangle }
{\langle \Psi_{GS} | \Psi_{GS} \rangle } = 
\frac{ \Gamma [ m - n + \frac{1}{2} ] \Gamma [ m - n + \frac{3}{2} ] }
{ 2 \pi N (M + 1) \Gamma^2 [ m - n + 1 ] }
\end{displaymath}

\beq
\times
\frac{ \Gamma [ m + 1 ] \Gamma [ M - m +
\frac{1}{2} ] }{ \Gamma [ m + \frac{3}{2} ] \Gamma [ M - m +1 ] }
\frac{ \Gamma [ n + \frac{1}{ 2} ] \Gamma [ M - n + 1 ] }{ \Gamma [ 
n + 1 ] \Gamma [ M - n + \frac{3}{2} ] } 
\label{bigas}
\eneq

\noindent
where $\langle \Psi_{GS} | \Psi_{GS} \rangle  = N^{M+1}
(2M + 2)!/2^{M+1}$ \cite{kwil}.

The definition of the wavefunction for two spinons in real space is now
straightforward. $\Psi_{\alpha \beta}$ is the state of two localized
spinons at $z_\alpha$ and $z_\beta$. Hence, we define the two-spinon
wavefunction, $z_\alpha^m z_\beta^n p_{mn} ( z_\alpha / z_\beta)$ from

\beq
\Psi_{\alpha \beta }  = 
\sum_{ m = 0}^M \sum_{n=0}^m (-1)^{m+n} z_\alpha^m z_\beta^n p_{ mn}
( \frac{ z_\alpha}{ z_\beta} ) \Phi_{mn}
\; \; \; .
\label{twospinon}
\eneq

\noindent
It is in principle possible to invert Eq.(\ref{tras1}) and to obtain
$p_{mn}$ algebraically.  However, we developed a much simpler approach,
which makes use of the fact that $\Psi_{\alpha \beta}$ is perfectly
defined for any $z_\alpha,z_\beta$ on the unit circle. Because $ | \Phi_{m
n} \rangle$ is an eigenstate of ${\cal H}_{HS}$, one obtains

\beq
\langle \Phi_{mn} | {\cal H}_{HS} | \Psi_{ \alpha \beta} \rangle = 
E_{mn} \langle \Phi_{mn} | \Psi_{ \alpha \beta} \rangle \; \; \; .
\label{dif1}
\eneq

\noindent
On the other hand, by standard manipulations \cite{chia,sutherland},
one can also show that

\begin{displaymath}
\langle \Phi_{mn} | {\cal H}_{HS} | \Psi_{ \alpha \beta}
\rangle = E_{GS} \langle \Phi_{mn} |  \Psi_{ \alpha \beta}
\rangle +
\end{displaymath} 

\begin{displaymath}
 \frac{J}{2} ( \frac{2 \pi}{N} )^2
\biggl\{ ( M - z_\alpha \frac{ \partial}{ \partial z_\alpha }) z_\alpha
\frac{ \partial}{ \partial z_\alpha} + ( M - z_\beta \frac{ \partial}{
\partial z_\beta} ) z_\beta \frac{ \partial}{ \partial z_\beta} 
\end{displaymath}

\beq
- \frac{1}{2}  \frac{ z_\alpha + z_\beta}{ z_\alpha - z_\beta} ( z_\alpha
\frac{ \partial}{ \partial z_\alpha} - z_\beta \frac{ \partial}{ \partial 
z_\beta}) \biggr\} \langle \Phi_{mn} | \Psi_{ \alpha \beta} \rangle
\; \; \; .
\label{differe}
\eneq

\noindent
Note the last term in this equation, which is the spinon interaction, is
{\it large} and diverges as one power of the spinon separation. Upon
equating Eq.(\ref{dif1}) to Eq.(\ref{differe}) we finally derive the
differential equation

\[
z ( 1 - z) \frac{ d^2 p_{mn}}{ d z^2} + \biggl[   \frac{1}{2} - m + n -
\]

\beq
( - m + n + \frac{3}{2} ) z \biggr] \frac{ d p_{mn}}{ d z} +
\frac{ m - n }{2} p_{mn} = 0 \; \; \; .
\label{grat}
\eneq

\noindent
The solution to Eq.(\ref{grat})  is the hypergeometric polynomial
\cite{abram}

\[
p_{mn} ( z ) = 
\frac{ \Gamma [ m - n + 1]}{ \Gamma [ \frac{1}{2} ] \Gamma [ m - n +
\frac{1}{2
} ] }
\]

\beq
\times
\sum_{ k = 0}^{ m-n} \frac{ \Gamma [ k + \frac{1}{2} ] \Gamma [
m - n - k + \frac{1}{2} ] }{ \Gamma [ k + 1 ] \Gamma [ m - n - k + 1] }
z^k \; \; \; .
\label{bigboy2}
\eneq

\noindent
In Fig.\ref{fig1} we plot $|p_{mn} ( z_\alpha / z_\beta )|$ vs. $\alpha -
\beta$. The sharp maximum at small spinon separation is a direct
consequence of the strong attractive interaction between the spinons
seen in Eq. (\ref{differe}).

We shall now prove rigorously that this enhancement is responsible for
the square-root singularity in the DDS.  The susceptibility is defined
by

\[
\chi_q(\omega) = \sum_X
\frac{| \langle X | S_q^- | \Psi_{GS} \rangle |^2}
{\langle X | X \rangle \; \langle \Psi_{GS} | \Psi_{GS} \rangle }
\]

\beq
\times
\frac{2 (E_X - E_{GS})}{(\omega + i \eta)^2 - (E_X - E_{GS} )^2}
\; \; \; ,
\eneq

\noindent
where $| X \rangle$ denotes an exact eigenstate of ${\cal H}$,
$E_{X}$ denotes its eigenvalue, and

\beq
S_q^- = \sum_\alpha (z_\alpha^* )^k (S_\alpha^x - i S_\alpha^y )
\; \; \; \; \; \; \;
(q = 2 \pi k / N ) \; \; \; .
\eneq

\noindent
However, since the act of flipping an $\uparrow$ spin to $\downarrow$
at site $\alpha$ is the same as creating two $\downarrow$ spinons on
top of each other at site $\alpha$ we have by virtue of Eq. 
(\ref{twospinon})

\[
S_q^- \Psi_{GS} = \sum_\alpha (z_\alpha^* )^k \Psi_{\alpha \alpha}
\]

\beq
= N \sum_{m = 0}^M \sum_{n = 0}^m (-1)^{m+n}  p_{mn}(1)
\delta(m+n-k) \; \Phi_{mn} \; \; \; .
\eneq

\noindent
Thus the set of two-spinon eigenstates exhaust the excited states
coupled to $\Psi_{GS}$ by $S_q^-$, and we have

\[
\chi_q (\omega) = N^2 \sum_{m = 0}^M \sum_{n = 0}^m
\frac{\langle \Phi_{mn} | \Phi_{mn} \rangle }
{ \langle \Psi_{GS} | \Psi_{GS} \rangle } p_{mn}^2 (1)
\]

\beq
\times \delta (m + n - k ) \;
\frac{2 (E_{mn} - E_{GS} )}{(\omega + i \eta)^2
- (E_{mn} - E_{GS} )^2 } \; \; \; .
\eneq

\noindent
This proves that the resonant enhancement is {\it entirely} due to
the functional form of $p_{mn}(z)$ shown in Fig. 1.

The thermodynamic limit is defined as $M\rightarrow \infty$, with $m/M$
and $n/M$ held constant. From general properties of the hypergeometric
functions \cite{abram} we obtain $p_{mn} ( 1 ) = \Gamma [1/2] \;
\Gamma [m - n +1] / \Gamma [m - n + 1/2]$.  Then approximating
all the gamma functions using Stirling's formula and converting the sums
on $n$ and $m$ to integrals over the 1-spinon Brillouin zone, we obtain
the Haldane-Zirnbauer formula for the DSS \cite{hazi}

\[
\chi_q ( \omega ) = \frac{J}{ 2 } \int_{ - \frac{ \pi}{2}
}^\frac{\pi}{
2} d q_1  \int_{ - \frac{  \pi}{2} }^{ q_1} 
d q_2\frac{ | q_1 - q_2 |\delta ( q_1 + q_2 - q  )
 }{ \sqrt{ E ( q_1 ) E ( q_2 
) } } 
\]

\beq
\times \frac{2 E(q_1 , q_2 ) }
{(\omega + i \eta)^2 - E^2 (q_1 , q_2 )}
\; \; \; ,
\label{ter}
\eneq

\noindent
where $E ( q )$ and $E ( q_1 , q_2 )$ are the one-spinon and the
two-spinon energies, respectively.  This may be exactly integrated
over $q_1$ and $q_2$, and the result is 

\[
\chi_q ( \omega ) =  \frac{J}{4}
\]

\beq
\times \frac{\Theta [ \omega_2 (q ) - \omega ] \;
\Theta [ \omega - \omega_{-1} ( q ) ] \;
[ \omega - \omega_{ +1} ( q )]}
{\sqrt{ \omega - \omega_{ -1} (q )}
\sqrt{ \omega - \omega_{ +1} ( q ) }} \; \; \; ,
\label{the2}
\eneq

\noindent
where $ \omega_{-1} ( q ) = (J/2) q ( \pi - q )$, $\omega_{+1} ( q ) =
(J/2) ( 2 \pi - q )( q - \pi )$ and $ \omega_2 ( q ) = (J/2) q ( 2 \pi - q
)$. 

We see that, in the thermodynamic limit, the resonant enhancement in
$p_{mn}$ turns into the square-root divergence in $\chi_q ( \omega )$ at
threshold. The origin of the branch cut is the threshold energy for the
creation of a spinon pair with total momentum $q$. The physical meaning of
this branch cut is that the spin wave is absolutely unstable versus decay
into a spinon pair. Hence, no sharp poles, corresponding to possible
low-energy spin-1 stable excitations, develop, but, on the contrary, the
spinon-pair threshold is the same as the spin-wave threshold. This last
observation points toward the main conclusion of our work: spinon
attraction is of fundamental importance for understanding relevant
low-energy properties of spin-1/2 antiferromagnets. It generates a
resonant enhancement of the probability for two spinons to be at the same
site. The resonant enhancement greatly increases the amplitude for a
spin-1 excitation to break into a spinon pair, on top of an uniform
two-spinon joint density of states. This effect is evident in the
thermodynamic limit of our formulas, where we show that the enhancement
turns into the branch cut in the DSS.

We wish to thank A. Tagliacozzo and D. I. Santiago for numerous useful
discussions. This work was supported primarily by the National Science
Foundation under grant No. DMR-9813899. Additional support was provided by
the U.S. Department of Energy under contract No. DE-AC03-76SF00515 and by
the Bing Foundation.

\end{document}